\documentclass{pasj00}
\draft

\begin{document}
\SetRunningHead{Awaki et al.}{Wide-band spectroscopy of Mrk 3 with Suzaku}
\Received{**/**/**}%{yyyy/mm/dd}
\Accepted{**/**/**}%{yyyy/mm/dd}

\title{Wide-band spectroscopy of the Compton thick Seyfert 2 galaxy Mrk 3 with Suzaku}

%%% begin:list of authors
\author{Hisamitsu \textsc{Awaki},\altaffilmark{1}
              Naohisa \textsc{Anabuki},\altaffilmark{2} 
              Yasuchi \textsc{Fukazawa},\altaffilmark{3} 
              Luigi \textsc{Gallo},\altaffilmark{4} 
              Shinya \textsc{Ikeda},\altaffilmark{1} \\
              Naoki \textsc{Isobe},\altaffilmark{5}
              Takeshi \textsc{Itoh},\altaffilmark{6}
              Hideyo \textsc{Kunieda},\altaffilmark{7}
              Kazuo \textsc{Makishima},\altaffilmark{5, 6} \\
              Alex G. \textsc{Markowitz},\altaffilmark{8,9} 
             Giovanni \textsc{Miniutti},\altaffilmark{10}
             Tsunefumi \textsc{Mizuno},\altaffilmark{3}
             Takashi \textsc{Okajima},\altaffilmark{8,9} \\
             Andrew \textsc{Ptak},\altaffilmark{8,9}
              James N. \textsc{Reeves},\altaffilmark{8,9,11}
              Tadayuki \textsc{Takahashi},\altaffilmark{12} \\
              Yuichi \textsc{Terashima},\altaffilmark{1} 
              Tahir \textsc{Yaqoob}\altaffilmark{8,9}
}
%%  \thanks{Example: Present Address is xxxxxxxxxx}}
\altaffiltext{1}{Department of Physics, Ehime University, Matsuyama, 790-8577, Japan}
\altaffiltext{2}{Department of Earth and Space Science,  Osaka University, \\
                        1-1 Machikane-yama, Toyonaka, Osaka 560-0043, Japan}
\altaffiltext{3}{Department of Physical Science, Hiroshima University, \\ 
                   1-3-1 Kagamiyama, Higashi-Hiroshima, Hiroshima 739-8526}
\altaffiltext{4}{SUPA, School of Physics and Astronomy, University of St Andrews,\\
                   North Haugh, St Andrews, KY16 9SS, UK}
\altaffiltext{5}{The Institute of Physical and Chemical Research (RIKEN), 2-1 Hirosawa, Wako, Saitama 351-0198}          
\altaffiltext{6}{Department of Physics, University of Tokyo, 7-3-1 Hongo, Bunkyo-ku, Tokyo}
\altaffiltext{7}{Department of Physics, Nagoya University, Chikusa-ku, Nagoya, 464-8602}
\altaffiltext{8}{Astrophysics Science Division, NASA Goddard Space Flight Center, \\
                     Greenbelt Rd. Greenbelt, MD 20771}
\altaffiltext{9}{Department of Physics and Astronomy, Johns Hopkins University, \\
                          3400 N. Charles St., Baltimore, MD 21218}
\altaffiltext{10}{Institute of Astronomy, University of Cambridge, Madingley Road, Cambridge, CB3 0HA, UK } 
\altaffiltext{11}{Astrophysics Group, School of Physical and Geographical Sciences, \\
                         Keele University, Keele, Staffordshire ST5 5BG, UK}        
\altaffiltext{12}{Institute of Space and Astronautical Science, Japan Aerospace Exploration Agency, \\
 3-1-1 Yoshino-dai, Sagamihara, Kanagawa 229-8510 Japan}
\email{awaki@astro.phys.sci.ehime-u.ac.jp}

%%% end:list of authors

%%% Please use the following style in case that sorting by 
%%% affilation is impossible. 
%
% \author{%
%   D-Firstname \textsc{D-Familyname}\altaffilmark{1}
%   E-Firstname \textsc{E-Familyname}\altaffilmark{1,2}
%   and
%   F-Firstname \textsc{F-Familyname}\altaffilmark{2}}
% \altaffiltext{1}{Address of Institute}
% \email{ddddd@xxx.xxx.xx.xx}
% \email{eeeee@xxx.xxx.xx.xx}
% \altaffiltext{2}{Address of Institute}

%% `\KeyWords{}' always has to be placed before `\maketitle'.
 %Do NOT move this preamble from here!

\maketitle

\begin{abstract}
We obtained a wide-band spectrum of the Compton-thick Seyfert 2 galaxy Mrk 3 with Suzaku.  The 
observed spectrum was 
clearly resolved into weak, soft power-law emission, a heavily absorbed power-law component, cold reflection, and many emission lines. 
The heavily absorbed component, absorbed by gas with a column density of 1.1$\times$10$^{24}$ cm$^{-2}$, has an intrinsic 2--10 keV luminosity of 
$\sim$1.6$\times$10$^{43}$ erg s$^{-1}$, and is considered to be direct emission from the Mrk 3 nucleus.
The reflection component was interpreted as reflection of the direct light off cold, thick material; the reflection fraction $R$ was
1.36$\pm$0.20.  The cold material is inferred to be located $>$ 1 pc from the central black hole of Mrk 3 due to the low 
ionization parameter of iron ($\xi$ $<$ 1 erg cm s$^{-1}$) and the narrow iron line width ($\sigma$ $<$ 22 eV). 
A Compton shoulder to the iron line was detected, but the intensity of the shoulder component was less than that expected from 
spherically distributed Compton-thick material. 
The weak, soft power-law emission is considered to be scattered light by ionized gas. 
The existence of many highly-ionized lines from O, Ne, Mg, Si, S, and Fe in the observed spectrum 
indicates that the ionized gas has a broad ionized structure, with $\xi$=10--1000.  
The scattering fraction with respect to the direct light 
was estimated to be 0.9$\pm$0.2\%, which indicates that the column density of the scattering region is about 3.6 $\times$
10$^{22}$ cm$^{-2}$. This high-quality spectrum obtained by Suzaku can be considered a template for studies of Seyfert 2 galaxies.
 
\end{abstract}

\section{Introduction}

Active galactic nuclei emit huge amounts of energy resulting from accretion of matter onto supermassive black holes.  
Strong hard X-ray emission from these nuclei is thought to be non-thermal, and represented by a power-law with a canonical photon index of $\sim$1.9.  
The strong emission irradiates material around the supermassive black hole, leading to reprocessing of the emission. 
Observing the reprocessed emission can reveal the structure of the nucleus, and indicate distribution, abundances, and 
ionization state of the circumnuclear material.  
One problem is that the reprocessed emission can easily be dominated by the strong nuclear emission, e.g., in Seyfert 1 nuclei.  
The X-ray emission from Seyfert 2 nuclei, however, is known to be attenuated by thick matter (e.g. Awaki et al. 1991).
Therefore, Seyfert 2 galaxies are among the best targets to explore the reprocessing environment around the central black hole.

Mrk 3 is an X-ray bright Seyfert 2 galaxy at a redshift $z$=0.013509 (Tifft, Cocke 1988).   
Ginga detected that direct light from the Seyfert 2 nucleus was heavily obscured by thick material with a column density ($N_{\rm H}$) of 
7$\times$10$^{23}$ cm$^{-2}$ 
(Awaki et al. 1990).  Its intrinsic luminosity in the 2--10 keV band was estimated to be 2$\times$10$^{43}$ erg s$^{-1}$, which was comparable to 
those of Seyfert 1 galaxies. Due to the obscuration of the direct light, soft X-ray emission with many highly-ionized lines 
was detected (e.g. Iwasawa et al. 1994, Sako et al. 2000).  A Chandra observation found that the soft X-ray emission 
was extended along the [OIII] ionization cone, and found that the ionized lines had signatures of emission from a 
warm, photoionized region (Sako et al. 2000).   The obscuration of the direct light has also allowed us to detect a reflection component 
in the hard X-rays of Mrk 3 (e.g. Turner et al. 1997, Cappi et al. 1999).  
The observed X-ray properties are consistent with Seyfert unification schemes (e.g. Antonucci, Miller 1985). 
From these previous observations, it is known that Mrk 3 has a complex spectrum, 
which consists of a weak power-law emission, heavily-absorbed power-law emission, cold reflection, 
and many emission lines, but it is difficult to resolve the X-ray spectrum into each component.
BeppoSAX demonstrated the power of broad-band observations by disentangling three of the components in Mrk 3: the soft X-ray emission, 
the heavily-absorbed component,  and cold reflection accompanied by an iron line (Cappi et al. 1999). 
High-quality spectra of Mrk 3 were obtained with XMM-Newton, and ionized gas and cold, thick material around 
the nucleus were investigated through measurements of highly-ionized lines and Fe K lines, respectively.   A Compton 
shoulder to the iron line due to Compton down-scattering gives us evidence of reflection by Compton-thick material (e.g. Matt 2002, Watanabe et al. 2003).  
The intensity ratio between Fe-K$\alpha$ and K-$\beta$ lines is a powerful tool to measure the ionization state of iron in the 
reflecting matter.  The observations by XMM-Newton indicated that iron line diagnostics have the potential to investigate the 
cold reflecting matter (Pounds, Page 2005; Bianchi et al. 2005).  

In this paper, we analyze a wide-band spectrum of Mrk 3 obtained by Suzaku (Mitsuda et al. 2007),
with the aims of resolving the spectrum into each component and estimating properties of the matter around the nucleus. 
Errors correspond to the 90\% confidence level for one interesting parameter
unless we explicitly note a different confidence level.

\section{Observation and Data Reduction}

Mrk 3 was observed with the Suzaku satellite on 2005 October 22--23 
during the SWG phase.
Suzaku has four X-ray telescopes (XRTs; Serlemitsos et al. 2007), with a spatial resolution of 2$^{\prime}$ 
(half power diameter). The focal plane detectors of the XRTs are X-ray CCD cameras,
which are referred to as an X-ray imaging spectrometer (XIS; Koyama et al. 2007). Three of the four XISs
(XIS0, 2, and 3) are front-illuminated CCDs (FI) and the other XIS (XIS1) is a back-illuminated CCD (BI).
The XISs are sensitive to 0.2--12 keV X-rays on a 18$^{\prime}\times18^{\prime}$ field of view.
XIS1 has enhanced soft X-ray sensitivity. Each XIS is equipped with 
two $^{55}$Fe calibration sources, which are illuminated at the CCD corners.  
Suzaku also has  a 
non-imaging hard X-ray detector (HXD; Takahashi et al. 2007).  The HXD has two types of detectors,
PIN and GSO, and yields sensitivity from $\sim$10 to $\sim$700 keV.
Mrk 3 was set on the XIS-nominal position, and XIS and HXD data were obtained simultaneously.  
The XISs were operated in the normal clocking mode. 
The observed XIS and HXD data were processed through the pipeline processing, provided 
as revision 1.2 data. We obtained cleaned event files of XIS and HXD, and then analyzed the 
data following nominal analysis procedures.

\subsection{XIS Reduction}
The net exposure of the cleaned event data was about 87 ks.  First we made X-ray images using the 
cleaned XIS data.  Figure 1 shows a 0.4--2 keV band image and a 2--10 keV band image. 
We found seven X-ray sources in the 0.4--2 keV band image.  
We corrected the astrometry of the Suzaku XIS image in adjusting the position of 
the brightest source in the 2--10 keV band image on the position of Mrk 3. 
Table 1 lists the detected sources' positions and possible candidates. 

We also list their average count rates, summed over the four XIS detectors in the 0.4--2 keV band. 
The observed 0.4--2 keV band X-ray flux of Mrk 3 was estimated to be 5.7$\times$10$^{-13}$ erg s$^{-1}$ cm$^{-2}$. 
The count rate of the faintest source thus corresponds to 5$\times$10$^{-15}$ erg s$^{-1}$ cm$^{-2}$ assuming 
that the source has the same spectral shape as Mrk 3 in the 0.4--2 keV band.  
There are bright regions in the corner of the hard band image. These regions were illuminated by the
$^{55}$Fe calibration sources.

%%%%%%%%%%%%%%%%%%%%%%%%  Table 1

We selected X-ray events within a 4.5$^{\prime}$ radius centered on Mrk 3 for the spectral analysis. 
There was an X-ray source (IXO 30) in the circular region (e.g. Colbert, Ptak 2002).  The source was too close to 
Mrk 3 to separate its X-ray emission from that of Mrk 3. Therefore, we included the 
X-ray events of this source when we extracted the X-ray events of Mrk 3.  Bianchi et al. (2005) found 
that the X-ray spectrum of IXO 30 obtained by XMM-Newton was represented by a power-law with a 
photon index of 1.77, and that IXO 30 was non-variable over a 2-year monitoring period.
Based on their result, the possible contamination from IXO 30 was estimated to be about 7\% of Mrk 3 in the 0.4--2 keV band.
In order to estimate an X-ray background for Mrk3, we accumulated  X-ray events excluding both 
the source region and regions irradiated by the calibration sources (see Figure 1). 
We obtained both the source spectra and background spectra of XIS-0, 1, 2, and 3. 

Response matrices (RMFs) and ancillary response files (ARFs) were generated for each XIS independently using 
{\it xisrmfgen} and {\it xissimarfgen} in HEASOFT 6.1.2 (Ishisaki et al. 2007). The ARF generator
takes into account the hydrocarbon contamination on the XIS optical blocking filters.  
To examine the accuracy of the XIS RMFs, we extracted spectra for the calibration sources, and then fit the 
spectra with a two-Gaussian model.
The best-fit center energies of the calibration sources were about 5 eV lower than the weighted values of 
the theoretical energies for the transitions of K$\alpha_1$ and K$\alpha_2$. The discrepancy of 5 eV
was within the accuracy ($\sim$0.2\% at Mn-K$\alpha$) of the energy calibration of the XIS. 
The line widths of the calibration sources were found to be 0 ($< $15) eV. 
Since the energy scale of each CCD was quite similar to within 4 eV,
we added the three spectra of the front-illuminated CCDs, XIS-0, 2, and 3 by using {\it mathpha}.
Figure 2a shows the summed XIS spectrum and the XIS-1 spectrum, both after background subtraction.  
The XIS team pointed out the existence of an artificial feature in the spectrum from 1.825 to 1.843 keV.
Thus we removed that energy range from the XIS spectra of Mrk 3.
In order to fit the summed spectrum, we also added RMFs and ARFs for XIS-0, 2, and 3 spectra into one response
file using {\it marfrmf} and {\it addrmf} in HEASOFT 6.1.2.

%%%%%%%%%%%%%%%%%%%%%%%%  Table 2

\subsection{HXD Reduction}
The HXD-PIN count rate of Mrk 3 was found to be about 0.15 ct s$^{-1}$, which was about 30\% of the internal (non-X-ray)
background (NXB) of the HXD-PIN. PIN background count rates are variable and strongly depend on 
the Suzaku satellite orbit (Kokubun et al. 2007). 
We extracted both the source  and background spectra with identical good time intervals.
An event file of the NXB, reproduced by a model of the NXB for the Mrk 3 observation,
was provided by the HXD instrument team.  The current accuracy of the PIN NXB model for a 1 day observation 
is deduced to be about 5\% (peak-to-peak residual), which roughly corresponds to 2\% at the 1$\sigma$ level.
Since the source data
was not corrected for instrument dead time, we performed the dead time correction by using {\it hxddtcor}.
The deadtime was 6.5\%, and the good time exposure came to be about 81 ks.

Since the HXD-PIN is not an imaging detector, we also had to subtract a contribution from the Cosmic X-ray Background (CXB).
A response matrix for a flat field was provided from the HXD instrument team.  We  simulated a CXB spectrum
observed by the HXD-PIN by assuming the following shape for the CXB: 9.0$\times$10$^{-9}$ (E/3keV)$^{-0.29}$ 
exp($-E$/40keV) erg cm$^{-2}$ s$^{-1}$ sr$^{-1}$ keV$^{-1}$ (Bolt 1986).
Then we added the simulated CXB spectrum to the NXB spectrum.
Figure 2a shows the PIN spectrum after the background subtraction.  The source flux in the 12--70 keV band was
9.4$\times$10$^{-11}$ erg  cm$^{-2}$ s$^{-1}$.

A response matrix for a point source at the XIS nominal position (ae\_hxd\_pinxinom\_20060814.rsp) 
was used for our spectral analysis.  A cross-normalization of the HXD-PIN relative to the XIS 
was derived using Suzaku observations of the Crab. We used the cross-normalization of 1.13
at the XIS nominal position (Ishida, the XRT team, 2006).

We did not use the HXD-GSO data in our analysis, because the background study of the HXD-GSO 
is in progress.  When the reproducibility of the GSO background is  improved, we will include 
GSO data in order to determine the shape of the X-ray spectrum in the high energy band. 

\section{Wide-band Spectrum}

A key advantage of the Suzaku observatory is its wide-band spectroscopy from sub-keV energies to above 100 keV.
We simultaneously fit XIS and HXD spectra with a baseline model based on previous observations. The 
baseline model is defined as follows: 

\begin{equation}
I({\rm ph} ~{\rm s}^{-1}~ {\rm cm}^{-2} ~{\rm keV}^{-1}) =e^{- \sigma_{a,1} N_{\rm H1}} [ PL1 + e^{- \sigma_{a,2} N_{\rm H2}}  PL2 + reflection +  ELs],
\end{equation}

\noindent
where $N_{\rm H1}$ and $N_{\rm H2}$ are column densities for power-law 
components PL1 and PL2, respectively.
The absorbing cross sections ($\sigma_{a,1}, \sigma_{a,2}$) are defined in the {\it phabs} model in XSPEC v11.3.  The abundances of Anders \& Grevesse (1989) were used.
For $\sigma_{a,2}$, the absorber was placed at the cosmological redshift of the source.
We fixed the high energy cutoffs of both PL1 and PL2 at 400 keV, because Cappi et al. (1999) obtained an lower limit 
to the cut-off energy ($E_{\rm cut}$) of the power law component $E_{\rm cut} >$200keV and our spectral fit could not constrain 
$E_{\rm cut} $.
The Compton reflection hump was reproduced by {\it pexrav}, keeping both the input photon index and $E_{\rm cut}$ equal to
those of PL2.  The cosine of inclination and iron abundance in {\it pexrav} were fixed at 0.45 and one times solar, respectively. The 
emission lines (ELs) seen in the spectrum were represented by the sum of Gaussian components:

\begin{equation}
ELs = \sum_{i} gauss (E_{\rm i},\sigma_{\rm i}, N_{\rm i}),
\end{equation}

\noindent
where $E_{\rm i}$, $\sigma_{\rm i}$, and $N_{\rm i}$ are the center energy, width, and intensity of the $i$th line.
For lines below 2 keV, it is difficult to set a tight constraint on line widths by a Suzaku observation.  Thus, we 
fixed them at the values obtained with the Chandra HETG (Sako et al. 2000).
The fixed values are much smaller than the energy resolution of the XIS (Koyama et al. 2007).
The best-fit value of $N_{\rm H1}$ was less than the Galactic 
column density 8.7$\times$ 10$^{20}$ cm$^{-2}$ (Stark et al. 1992), thus we fixed the  $N_{\rm H1}$  at the Galactic value.

The observed spectra were fit with the baseline model in the energy range from 0.4 keV to 
70 keV, yielding a reduced $\chi^{2}$ of 1.15 (d.o.f.=762).  Figure 3 shows X-ray spectra in the 0.4--3 keV and 5.5--7.5 keV bands.
Many emission lines are detected.
The best-fit parameters of the continuum emission components are listed in Table 2, and those of the 
emission lines are summarized in Table 3.  An unfolded broad-band spectrum and the  baseline model are shown 
in Figure 2b. The baseline model is the same as the best-fit model except only that  the line width of He-like Fe 
was the same as that of the Fe K$\alpha$ line.  
We included two lines for Si XIII and Si XIV in the baseline model, 
and obtained consistent results with those of XMM-Newton,  
but the result may have some ambiguity due to a uncertainties with the response function around the Si edge. 
We obtained a best-fit photon index and the absorbing column density of 1.80$\pm$0.06
and (1.10$\pm$0.06)$\times$10$^{24}$ cm$^{-2}$, respectively. 

The observed 2--10 keV flux was calculated to be 6.4$\times$10$^{-12}$ erg s$^{-1}$ cm$^{-2}$, which
is similar to those of the BeppoSAX (6.5$\times$10$^{-12}$ erg s$^{-1}$ cm$^{-2}$) and XMM-Newton 
(5.9$\times$10$^{-12}$ erg s$^{-1}$ cm$^{-2}$) observations,  though it is slightly different from the Ginga
result ( 8.8$\times$10$^{-12}$ erg s$^{-1}$ cm$^{-2}$).  
The larger flux in the Ginga observation compared to the BeppoSAX and XMM-Newton observations 
is due to the difference in the absorption column.
The intrinsic X-ray luminosity of the heavily absorbed  component was estimated to be about 
1.6$\times$10$^{43}$ erg s$^{-1}$ in the 2--10 keV band, assuming $H_{\rm 0}$ = 70 
km s$^{-1}$ Mpc$^{-1}$ and $\Lambda_{\rm 0}$ = 0.73. 
The X-ray luminosities of the other components were estimated to be
1.4$\times$10$^{42}$ erg s$^{-1}$ in the 2--10 keV band for the reflection component and 3.7$\times$10$^{41}$ erg s$^{-1}$ 
in the 0.4--2 keV band for the PL1 component + ELs.  

The reflection ratio $R$ of the cold reflection to PL2 was about 1.36.  To estimate the error on $R$,
we have to consider the current systematic uncertainties in the background subtraction and the relative PIN/XIS 
cross-normalization.  The statistical uncertainty of the relative PIN/XIS cross-normalisation is only about 3\% (Ishida, the XRT team 2006).
The change in $R$ caused by this uncertainty is smaller than that from the uncertainty of the background subtraction.
Thus, we evaluated the effect of the NXB reproducibility on $R$. 
We modified the intensity of the PIN non X-ray background by $\pm$2\%, which is about the 1$\sigma$
 level of the current reproducibility of the PIN NXB for one day exposure in the 15--40 keV band
 (Figure 3 in Mizuno et al. 2006). The net count rates of the source were changed to be
+15\% and --6\% for the --2\% and +2\% NXB, respectively. The best-fit value of $R$ became 1.25 and 1.42, respectively.
This change is smaller than the statistical error on $R$, $\pm$0.20, determined by the fitting procedure.  
The value of  $R$ derived from the Suzaku observation  is different from the BeppoSAX 
value ($R$ = 0.95$^{+0.12}_{-0.15}$).  

To estimate the scattering 
fraction, we have to subtract the contamination from IXO 30.  Since the 2--10 keV flux of IXO 30, as 
observed by
XMM-Newton, was $\sim$8.5$\times$10$^{-14}$ erg cm$^{-2}$ s$^{-1}$ (Bianchi et al. 2005),  the contamination 
of IXO 30 into PL1 was estimated to be about 17\%. 
Thus the ratio of PL1 with respect to PL2 was found to be 0.9$\pm$0.2\%.

%%%%%%%%%%%%%%%%%%%%%%%  Table 2
We compared the center energies of ELs with those obtained by XMM-Newton and Chandra, and found
that the center energies were consistent with those values within 20 eV. 
We also found that our line intensities were more comparable to those for XMM-Newton rather than Chandra. 
Figure 4 shows the comparison between our result  and the Chandra result. The difference is
larger in the soft band and becomes unity above 2 keV. Sako et al. (2000) pointed out the spatial extent of the soft band image,
which laid approximately in the cross dispersion direction of the
HETGS.  Since Sako et al. (2000) applied a 8 pixel filter in order to reduce the noise level in the dispersed spectrum,  their spectrum did not 
cover all the extended emission. 
The consistency with the XMM-Newton results can be explained by the fact that the line-emitting region is extended. 
The equivalent widths of these lines to the PL1 and to the reflection are listed in Table 4.
For the iron line, the line width of about 20 eV ($<$ 29 eV) and the intensity of about 4.7 $\times$10$^{-5}$ ph 
s$^{-1}$ cm$^{-2}$  are more consistent with the Chandra grating results rather than the XMM-Newton result 
(Bianchi et al. 2005).  
The equivalent width of the iron line to the reflection component was estimated to be $\sim$855 eV.  

\subsection{\textit{A Compton Shoulder (CS)}}
A Compton shoulder component is often represented by a second Gaussian component with a width of $\sim$80 eV at 
an energy about 0.1 keV lower than the primary Fe K$\alpha$ Gaussian (e.g. Matt 2002). 
However, we represented the component by a pulse function with a pulse width of 156 eV in this paper, 
since the shape of the shoulder component is more similar to a pulse function rather than a Gaussian function 
(e.g. Matt 2002, Watanabe et al. 2003). Furthermore, the pulse width of 156 eV corresponds to the maximum energy shift 
due to a single Compton scattering.  By fixing the pulse width at 156 eV in our fitting procedure, the CS can be represented by
only one parameter, $f$, which is the fraction of the CS intensity with respect to that of the primary Gaussian component.
We fit the iron line bandpass of Mrk 3 with the Gaussian plus pulse components, finding that
the pulse component was detected 
at the 90\% confidence level. In this fitting procedure,  both the parameters of many lines except of the 
iron K$\alpha$ line and the photon index were fixed at the best-fit values.  The best-fit ratio $f$ of the pulse component to
the Gaussian component was 10$\pm$8\%.  

When including the CS, the best-fit values of the line center energy and the line width 
moved slightly to 6.406 keV and 0 ($<$22) eV, respectively.  The line width of $\sigma$ $<$ 22 eV corresponds to a 
velocity of $<$ 1000 km s$^{-1}$ due to Doppler broadening.
The line width of 20 eV in the baseline model 
can be explained by the existence of the CS.  The effects of the CS to the best-fit center energy and width 
are shown in a $\chi^{2}$ contour map (Figure 5).

\subsection{\textit{Iron K$\alpha$ and K$\beta$ Lines}}
By determining the center energies and intensities of the iron K$\alpha$ and K$\beta$ lines accurately, we can 
measure the ionization state of the line-emitting gas (e.g. Bianchi et al. 2005, Yaqoob et al.\ 2007).
At first, we obtained a line intensity ratio $I_{K\beta}$/$I_{K\alpha}$  of 0.10$\pm$0.03 in the baseline model (no CS model). 
The ratio was determined more accurately than obtained by Bianchi et al. (2005). 
In comparison with a calculation of the line ratio in the low ionization region of Fe by Palmeri et al.
(2003), we constrained the ionization state 
to Fe I--IX.
Figure 6 shows the confidence contours of the center energies of K$\alpha$ versus K$_{\beta}$.  
The calibration uncertainty of the XIS response at 6 keV is about 15 eV (Koyama et al. 2007).
The center energies of the iron K$\alpha$ and K$\beta$ lines may be shifted within the calibration uncertainty. 
We shifted the contour map by $\pm$7.5 eV.  If the contour maps is shifted by --3 eV, the filled circles of
Fe II to Fe X are included in the 90\% confidence region. The line ratio and the line energies indicate a low
ionization state of the reflection matter.
As noted above, the central energy of the iron K$\alpha$ line shifted from 6.401 keV to  6.406 keV  
when the CS was included into the iron K$\alpha$ profile.

\subsection{\textit{Advanced Analysis of the Reflection Component}}

A Compton reflection component with a reflection factor $R$ of $\sim$1.36 was clearly detected.  Although 
we assumed one solar
abundance of iron in the baseline model, we can measure the iron abundance ($A_{\rm Fe}^{\rm ref}$) from the edge seen in the 
reflection spectrum. We made $A_{\rm Fe}^{\rm ref}$ of {\it pexrav} free in our fitting procedure, and obtained the best-fit value of 
$A_{\rm Fe}^{\rm ref}$ = 0.56$\pm$0.10 with an improvement of $\Delta\chi^{2}$ = --32.   
Because an edge structure was also produced by the heavy absorption with $N_{\rm H2}$, the best-fit $A_{\rm Fe}^{\rm ref}$
was coupled with that ($A_{\rm Fe}^{\rm abs}$) of the absorption matter. 
The  dependence of these abundances is shown in a $\chi^{2}$ confidence map (see Figure 7).
If the reflecting matter is identical to the absorbing matter,  $A_{\rm Fe}^{\rm ref}$ should be similar to be 
$A_{\rm Fe}^{\rm abs}$.  Therefore, we estimated the iron abundance assuming $A_{\rm Fe}^{\rm ref}$
= $A_{\rm Fe}^{\rm abs}$. The abundance was found to be 0.84$\pm$0.05 with $\Delta\chi^{2}$ = --28.

We measured the energy of the Fe edge seen in the reflection component.  Since there is no parameter on 
 the edge energy in {\it pexrav}, we changed the parameter {\it redshift} in the {\it pexrav} model in order to obtain the 
 confidence region of the edge energy. The best-fit 
value of the redshift was 0.014$^{+0.007}_{-0.009}$, which corresponds to a confidence region of the edge energy of 
7.11$^{+0.06}_{-0.05}$ keV after correcting for the cosmological redshift.  

We examined the possibility of obscuration of the reflection component in the baseline model, because if the covering 
factor of the reflecting matter is large, the reflection component may be obscured by itself.  Including the absorption $N_{\rm H}^{\rm ref}$ 
of the reflection component in the baseline model, the fit was significantly improved, with $\Delta \chi^{2}$ $\sim$  --7.  
The $F$-test probability for the improvement of $\chi^{2}$ is estimated to be about 1\%.
The best-fit photon index changed to 1.83 in the case of $N_{\rm H}^{\rm ref}$ =1.7$\times$10$^{22}$ cm$^{-2}$.
If the iron abundance was changed to be subsolar instead of one times solar, obscuration of the reflection component was not 
required ($N_{\rm H}^{\rm ref}$ $<$ 0.9 $\times$10$^{22}$ cm$^{-2}$).
 
\subsection{\textit{A Highly-Ionized Iron Line}}
A highly-ionized iron line at the energy of 6.682$^{+0.021}_{-0.023}$ keV is detected at the $\sim$3$\sigma$ level. 
This energy is consistent with the energy of a resonance ($E$ = 6.700 keV) or an intercombination line ($E$ = 6.675 keV) rather than
a forbidden line ($E$ = 6.637 keV). The intensity of this line was found to be 3.9$^{+1.5}_{-1.2}$$\times$10$^{-6}$ ph cm$^{-2}$ s$^{-1}$.
The observed equivalent width with respect to the scattered light (PL1) was about 700 eV. Netzer et al. (1998) estimated the equivalent width of 
a highly-ionized iron line to the scattered light. Our observed equivalent width is consistent with their estimation of 500--1000 eV.  

We examined the existence of the hydrogen-like iron line at $E$ = 6.966 keV in the rest frame, but we obtained only an upper limit of 
2$\times$10$^{-6}$ ph 
 cm$^{-2}$ s$^{-1}$.  

\subsection{\textit{Absorption model for $N_{\rm H}$ $>$ 10$^{24}$ cm$^{-2}$}}
The PL2 source  is heavily obscured by thick matter with $N_{\rm H2}$=1.1$\times$10$^{24}$ cm$^{-2}$.  
For $N_{\rm H}$ $>$ 10$^{24}$ cm$^{-2}$, the effect of the Compton scattering have to 
be considered,
because optical depth of the Compton scattering is close to unity  (Ghisellini et al. 1994, Yaqoob 1997). 
Since the $phabs$ in XSPEC considers only photo-electric absorption, we tried to use $plcabs$ in XSPEC which 
describes X-ray transmission of an isotropic source of photons located at the center of a uniform, spherical distribution 
of matter, correctly taking into account Compton scattering.
This model is not valid above 18 keV, but it is useful to know the effect of the Compton scattering.

The best-fit parameters of the continuum components are listed  in Table 5.   The fit with $plcabs$ is statistically better than 
the our baseline model ($\Delta\chi^{2} \sim -10$).  This model gives the best fit column density $N_{\rm H2}$ of
1.1$\times$10$^{24}$ cm$^{-2}$ which is consistent with that obtained with the baseline model, and a steep photon
index of 1.88. The reflection ratio and the scattering ratio are changed to $R$$\sim$1.1 and $C_{\rm f}$$\sim$0.6\%
due to the increase of the intensity of the direct component.  The intrinsic luminosity strongly depends on the best fit 
model. The 2--10 keV intrinsic luminosity with $plcabs$ model is
estimated to be  2.2$\times$10$^{43}$ erg s$^{-1}$, which is larger than that obtained by the baseline model. 

For the emission lines, most of the best parameters do not significantly change.  The best-fit center energy of Fe K$\beta$ 
line reduced to be 7.04$\pm$0.03 with the intensity of 3.9$\times$10$^{-6}$ ph cm$^{-2}$ s$^{-1}$. 
The energy of the K$\beta$ line is still consistent with that for neutral Fe.

\section{\textit{Timing Analysis}}
Thanks to Suzaku's wide bandpass, we detected a direct emission component from Mrk 3.
We made a light curve in the 15--40 keV band with the Suzaku PIN to examine the time variability of the direct component, which is
dominant in this energy band. Background subtraction and deadtime correction were performed for the light curve.
The light curve is shown in Figure 8. The systematic error of the non X-ray background in the 15--40 keV band for a 5760 s bin was 
about 6\% (1$\sigma$) of the non X-ray background  (Mizuno et al. 2006). Considering this systematic error, we found that the light 
curve was stable during the observations, with a reduced 
$\chi^2$ of 0.40 (d.o.f.=32) against a constant model.  The 90\% upper limit to the time variability was estimated to be 
$\sigma^{2}_{\rm rms}$$<$0.01.

We examined the long-term variability of Mrk 3 by comparing the Suzaku result with the BeppoSAX result, obtained during 1997 April 16--18. 
The 2--10 keV intrinsic luminosity observed by BeppoSAX can be estimated from the 0.1--150 keV
luminosity by Cappi et al. (1999).   Although the derived luminosity depends on the best fit model 
(e.g., it depends on values of $\Gamma$, $N_{\rm H}$, and $R$), we obtained an
X-ray luminosity of 2.7$\times$10$^{43}$ erg s$^{-1}$.
We also compared the Ginga and XMM-Newton results, although these results have some ambiguity
due to being narrow band observations. The Ginga luminosity during the period of 
1989 September 27--28 and the XMM-Newton luminosity during 2000
October 19--20 were estimated to be 2.3$\times$10$^{43}$ erg s$^{-1}$ 
(e.g. Awaki et al. 1991, Griffiths et al. 1998) and 2.2$\times$10$^{43}$ erg s$^{-1}$
(e.g. Bianchi et al. 2005), respectively.   Long-term variability of Mrk 3 was  pointed out by Cappi et al. (1999). 
There may be a gradual decrease in the intrinsic luminosity between the BeppoSAX, XMM-Newton
and Suzaku observations. Further observations with Suzaku are crucial to quantifying the long term
variability of the intrinsic luminosity of Mrk 3.

\section{Discussion}

\subsection{\textit{Continuum Emission}}
We obtained a wide-band spectrum covering energies from 0.4 to 70 keV.
The continuum emission was resolved into three components: a soft, weak power-law, a cold-reflection component,
and a heavily absorbed power-law component.   We reproduced well the spectrum assuming the 
same photon index for each component.  

The heavy absorbed component with $N_{\rm H2} \sim$ 1.1$\times$10$^{24}$ cm$^{-2}$ is considered to be 
a direct component  from the nucleus, since the photon index of $\sim$1.8 is similar to the canonical value for Seyfert 1 
galaxies and the intrinsic X-ray luminosity in the 2--10 keV band was estimated to be 1.6$\times$10$^{43}$ erg s$^{-1}$.
The column density we obtained is consistent with the BeppoSAX and XMM-Newton observations:
For the last decade, Mrk 3 has remained a Compton-thick object.
Some time variability of the absorbed component is expected, similar to Seyfert 1 galaxies (e.g. Markowitz et al. 2003).  
Mrk 3's intrinsic luminosity has evolved somewhat over the last decade.  However, short-term time variability
was not significantly detected over the 2 day observation.  The black hole mass ($M_{\rm BH}$) has been estimated to be 
4.5$\times$10$^8$ $M_{\odot}$ from optical observations (Woo, Urry 2002).  $\sigma_{\rm rms}^{2}$ for a 2 day 
observation was expected to be $\sim$3$\times$10$^{-4}$ from the relation between $M_{\rm BH}$ and $\sigma_{\rm rms}^{2}$ 
(Papadakis 2004); this is consistent with the upper limit obtained during the Suzaku observation.

\subsection{\textit{Cold Reflection}}

The cold reflection component was detected, with a reflection factor $R$=1.36$\pm$0.20, and the
observed X-ray flux in the 2--10 keV band was (3.0$\pm$0.1)$\times$10$^{-12}$ erg s$^{-1}$ cm$^{-2}$ in
the baseline model.   
The value of $R$ is different from that from the BeppoSAX
observation, 0.95$^{+0.12}_{-0.15}$, and the
observed flux in the 2--10 keV band
is consistent with the X-ray flux of $\sim$ 3.4$\times$10$^{-12}$
erg cm$^{-2}$ s$^{-1}$ measured during the XMM-Newton
observation (Bianchi et al. 2005). The direct nuclear emission
may be gradually decreasing from the time of the BeppoSAX
observation, to the XMM-Newton and Suzaku observations. If
the reflector is a pc scale torus, then a time delay between the
reflection component and the direct component is likely to be
on a timescale of years. Continued X-ray spectral monitoring
of Mrk 3 with Suzaku would thus be useful for constraining
the location of the reprocessing matter.

We detected the CS of the iron K$\alpha$ line with a 90\% confidence level. It is known that the intensity ratio $f$ 
 depends on the column density of the scattering/reflecting material (e.g. Matt 2002, Watanabe et al. 2003). 
The value of $f$ = 0.1 roughly corresponds to an optical depth of Compton scattering: $\tau \sim \sigma_{\rm T}$ $N_{\rm H}^{\rm cs}$,
where $\sigma_{\rm T}$ and $N_{\rm H}^{\rm cs}$ are the Thomson cross-section and the column density of the reflecting matter, respectively.
$N_{\rm H}^{\rm cs}$ was estimated to be 2 $\times$10$^{23}$ cm$^{-2}$, which is smaller than the absorbing 
column $N_{\rm H2}$.  This means that a uniform density spherical distribution of reflecting matter around an X-ray emitter
was not appropriate.   Matt (2002) calculated the CS properties for an isotropically illuminated plane-parallel slab.
In this geometry, $f$ is smaller than that in a spherical distribution.  In order to explain a small value of $f$, we have to consider 
a geometric configuration wherein material has a large optical depth before scattering or absorption by iron, and a small optical depth 
for reprocessed light.  A torus geometry is probably worth studying for at least part of the Compton shoulder emission. 

The ionization parameter, $\xi$, of the reflecting matter can be estimated from the iron K$\alpha$ and K$\beta$ lines.  
From the observed intensity ratio of these lines,  a low ionization state, Fe I--IX, has already been pointed out.  We examined the low ionization state
using the center energies of these lines. 
Palmeri et al. (2003) calculated theoretical energies of the Fe K$\alpha$ and K$\beta$ lines for  Fe$_{\rm II}$--Fe$_{\rm IX}$.  
 We plotted the theoretical energies on Figure 6. The plot also demonstrates that  the ionization state of iron is lower than Fe IX.  
 The low ionization state of iron was also confirmed with the
iron edge energy.  
Kallman, Baustina (2001) and Kallman et al. (2004) estimated ion  
fraction of a low-density ionized  gas as a function of ionization parameter $\xi$ . Their estimate  
is  valid for $N_{\rm H}$ $<$ 10$^{24}$ cm$^{-2}$.
  We suppose that the absorber is the same as the reflector. In the  
case of a pc scale torus, the
  mean density of the torus is estimated to be 3$\times$10$^{6}$ cm$^{-3}$.  The  
ionization parameter with the
  largest emissivity for Fe IX is estimated to be about 1 erg cm s$^{-1} $.
 The luminosity ($L^{\rm ion}$) to be used when estimating the ionization parameter is defined as the luminosity in the 1 -- 1000 ryd band. 
Assuming a simple power-law
model with an energy index of 0.8, $L^{\rm ion}$ was deduced to be 3 times the luminosity in the 2--10 keV band.
From the definition of $\xi$, the distance $r$ from an X-ray source with $L^{\rm ion}$ was deduced as follows:

\begin{equation}
r = ( \frac{L^{\rm ion}}{n  \xi} )^{1/2}\sim 2.3 (\frac{L^{\rm ion}}{5\times10^{43} {\rm ~erg~s^{-1}}})^{1/2} (\frac{n}{10^{6} {\rm ~cm^{-3}}})^{-1/2} (\frac{\xi}{1 {\rm ~erg~cm~s^{-1}}})^{-1/2}  (\rm pc)
\end{equation}

\noindent
The ionization parameter indicates that the reflecting region is located at $r$  $>$ 1 pc from the X-ray emitting region.

The location of the reflecting region was also estimated from the iron line width, assuming Keplerian motion of the iron line-emitting matter
and Doppler broadening of the iron line.  The velocity ($v_{\rm FWHM}$) estimated from the line width is expected to be about 2$v_{\rm k} \sin i$,
where  $v_{\rm k}$ is the Keplerian velocity and $i$ is the angle between the rotation axis and our line of sight.  
The distance $r$ from the center of gravity is deduced as follows:

\begin{equation}
r \sim ( \frac{v_{\rm FWHM}}{1300 {\rm~km~s^{-1}} } )^{-2} M_{8} \sin^{2} i  (\rm pc),
\end{equation}

\noindent
where M$_{8}$  is the total amount of mass within $r$ in units of 10$^{8}$ $M_{\odot}$.   Assuming that the central black hole mass dominates 
the total mass, the central black hole mass of 4.5$\times$10$^{8}$ $M_{\odot}$ and the upper limit to the line width of 2400 km s$^{-1}$ 
yield an estimate on the distance: r $>$ 1.3 $\sin^{2} i$ pc.  Assuming that the inclination angle $i$ is equal to the narrow line region
(NLR) inclination angle of 80--85$^{\circ}$ (Ruiz et al. 2001), the distance is greater than 1.3 pc, which corresponds to
$\sim$3$\times$10$^{4}$ 
$r_{\rm s}$, where $r_{\rm s}$ is the Schwarzschild radius for a black hole mass of 4.5$\times$10$^{8}$ $M_{\odot}$.  
The estimate using the iron line width also suggests that the reflecting region is far from the central black hole.

The iron abundance of the reflecting region was estimated from the
iron edge structure seen in the reflected component, and an abundance
of $\sim$0.56 was obtained for cos$i$ = 0.45. Linking it with the
abundance of the absorbing column density $N_{\rm H2}$, the abundance
changed to $\sim$0.84 solar. Metal abundances were also estimated from
the equivalent width of the iron K$\alpha$ line with respect to the
reflection component.  For $\cos i$=0.45 and a metal abundance of one
solar, the equivalent width of an Fe K$\alpha$ line with respect to a
reflection component is derived to be about 1200 eV assuming a plane-parallel,
semi-infinite slab of neutral matter (e.g., Matt et al. 1997, Pounds \&
Page 2005). Matt et al. (1997) used the Anders \& Grevesse abundances
(hereafter AG abundances), which is the same as what we used in XSPEC.
The observed equivalent width of 855$\pm$35 eV corresponds to an iron
abundance of 0.7$\pm$0.1 solar, which is consistent with the iron abundance
estimated from the edge structure. The iron abundance in Mrk 3 is thus
estimated to be sub-solar, using the AG abundances and the plane-parallel
slab geometry.  We compared the abundance with those of other AGNs observed
by Suzaku. The iron abundances in  MCG--5-23-16 and Cen A were measured
to be 0.65$\pm$0.15  solar (Reeves et al. 2007) and 1.2$\pm$0.1 solar
(Markowitz et al. 2007), respectively.  However, since those authors
used different abundance tables (Wilms, Allen, \& McCray (2000) and
Lodders (2003), respectively), we converted them to the iron
abundances in the case of the AG abundances. The converted abundances
were $\sim$0.4 and $\sim$0.8 solar in MCG--5-23-16 and Cen A, respectively.
The iron abundance in Mrk 3 is therefore similar to those of these AGNs;
taking the differences between various abundance table into account, the
iron abundance of Mrk 3 ranges from 0.5 to $\sim$1 solar.

We estimated metal abundances of other elements from equivalent widths
of those lines with respect to the reflection component. Matt et al.
(1997) also calculated the equivalent widths of Si and S K$\alpha$
lines to be 240 eV and 165 eV for cos$i$=0.45 and a metal abundance
of one solar, respectively.  Comparing these predictions to the
observed equivalent widths of 300$\pm$35 eV and 160$\pm$30 eV,
respectively,  for these lines, the metal abundances of  Si and S
are estimated to be about  one solar.  We note that there is some
ambiguity for the observed intensity of the Si K$\alpha$ line, but
the intensity is consistent with that obtained using XMM-Newton
(see Table 3). If the Si and S K$\alpha$ lines are emitted at the
reflection region along with the Fe lines, the abundance ratios of
$Z_{\rm Si}$/$Z_{\rm Fe}$ and $Z_{\rm S}$/$Z_{\rm Fe}$ are
1.7$\pm$0.2 and 1.3$\pm$0.2, respectively.  These results are similar
to those obtained for Cen A by Markowitz et al. (2007). The abundance
ratios suggest that the metals may be produced by type II SNe (e.g.
Tsuru et al. 2007). However, we have to consider the possibility
of contamination of the Si K and S K lines from emission due to
a photoionized region with a low ionization parameter, since the
photoionized region has a broad ionization structure, as discussed
below. This possibility could be related to the difference in
observed Si K$\alpha$ intensities between the Suzaku and Chandra
observations.

\subsection{\textit{The Scattering Region}}

The soft continuum emission, described by the power-law component PL1, is considered to be scattered light from the
photoionized region, as with polarized light in the optical band.
Thanks to Suzaku's wide bandpass, the ratio $C_{f}$ of the scattered to direct components
is accurately estimated to be 0.9$\pm$0.2\%.  This ratio is consistent with that assumed by Miller \& Goodrich (1990).
The ratio approximately corresponds to  $\tau$ $\Delta \Omega$/4$\pi$, where
$\Delta \Omega$ is the covering factor of the scattering region, and  $\tau$ is the scattering optical depth. The column density 
($N_{\rm H}^{\rm scat}$) for the scattering is described as $\tau$/$\sigma_{\rm T}$, where $\sigma_{\rm T}$ is Thomson cross-section.

\begin{equation}
N_{\rm H}^{\rm scat} =  \frac{\tau}{\sigma_{\rm T}} = 1.3\times 10^{22} (\frac{C_{f}}{0.01}) (\frac{\Delta \Omega}{4\pi})^{-1} {\rm cm}^{-2}
\end{equation}

\noindent
The opening angle of the scattering region may be estimated from the opening angle of the NLR. Capetti et al. (1995) found a
NLR opening angle of $>$100$^{\circ}$.  
Assuming a half opening angle 50$^{\circ}$ for the bipolar ionization cone (Capetti et al. 1995), $N_{\rm H}^{\rm scat}$ is 
deduced to be 3.6$\times 10^{22}$ cm$^{-2}$. 
The half opening angle of 50$^{\circ}$ is larger than the estimate by Ruiz et al. 
(2001) due to the inclusion of all the Z-shape emission components in the NLR in the estimates of 
Capetti et al. (1995). 
PL1 had an identical photon index to PL2, and had a low absorption column 
corresponding to $N_{\rm H} \sim$ 6.4$\times$10$^{20}$ cm$^{-2}$. This suggests that the scattering region is highly ionized.  
In the case of a uniform electron density, the size of the
scattering region is estimated to be about 50 ($L$/5$\times$10$^{43}$ erg s$^{-1}$) ($N_{\rm H}$/3.6$\times$10$^{22}$ cm$^{-2}$)$^{-1}$ 
($\xi$/10)$^{-1}$ pc, and the density in the scattering region is estimated to be about 240 cm$^{-3}$.

Information on the ionized region can be obtained from the emission-line spectrum.  A photoionized plasma
can be characterized by an ionization parameter $\xi$. 
Since He-like iron and He-like oxygen do not coexist in a photoionized plasma with a uniform $\xi$ (Kallman, Bautista 2001), the ionized gas in Mrk 3 must have a broad ionization structure with $\xi$=10--1000, as 
pointed out by previous observations (e.g. Pounds, Page 2005).

 As mentioned above, the scattering region may extend up to 1 kpc (Sako et al. 2000). This is inconsistent with the estimate
 of the uniform electron density model.  We tried to explain the extent of the scattering region by assuming that the density gradually 
 decreases as the distance from the nucleus increases, i.e. $n$=$n_{\rm 0} r^{-1}$. The column density was derived from the integral 
 of the density from $r$=$r_{\rm in}$ to $r_{\rm out}$, where 
 $r_{\rm in}$ and $r_{\rm out}$ are the inner and outer radii of the scattering region, respectively. In $r$ $<$ 1 pc, the scattering region might be
 obscured by the dusty torus. We thus set $r_{\rm in}$=1pc. For $r_{\rm out}$, we adopted the extent observed by Chandra.
For  $N_{\rm H}$=3.6$\times$10$^{22}$ cm$^{-2}$, the electron densities at 1 pc and 1 kpc were estimated to be 
1700 cm$^{-3}$ and 1.7 cm$^{-3}$, respectively. The ionization parameters $\xi$ at 1 pc and 1 kpc are 3300 and 3.3, respectively.
This is consistent with the broad ionization structure.

\section{Summary and Conclusion}
We obtained a wide-band spectrum of Mrk 3 with Suzaku. The spectrum in the 0.4--70 keV band was described by our baseline
model, which consists of a power-law emission component, heavily obscured power-law emission, a cold reflection component, and many emission lines.
The heavily absorbed component is considered to be a direct nuclear emission component. The photon index of the obscured component was 
1.80$\pm$0.06.
The intrinsic luminosity 
of the direct component was estimated to be 1.6$\times$10$^{43}$ erg s$^{-1}$ in the 2--10 keV band. We did not find any significant time 
variability during the 2-day observation, but there is evidence for long-term variability from a comparison with results of previous observations.
The nuclear activity in the 2--10 keV band is almost completely obscured by thick matter with $N_{\rm H2}$ $\sim$1.1$\times$
10$^{24}$ cm$^{-2}$, which suggests that Mrk 3 was in a Compton-thick phase. 

The cold reflection component was detected with a reflection factor of $R \sim$1.36. The observed luminosity was  estimated to be
1.4$\times$10$^{42}$ erg s$^{-1}$ in the 2--10 keV band. 
A strong iron K$\alpha$ line was detected with the EW of $\sim$855 eV with respect to the cold reflection.  
We examined the existence of  a Compton shoulder to the iron K$\alpha$ line, and detected the shoulder with 90\% confidence. 
However, the intensity of the shoulder component
 was smaller than that expected from Compton reflection by spherically distributed thick matter. To resolve this issue, we need to 
 consider a suitable distribution for the thick matter. 
We evaluated an ionization parameter $\xi$ of the reflection component by analyzing iron K$\alpha$  and K$\beta$ lines in detail. 
The ionization state of iron was determined to be lower than Fe IX. This was also confirmed from the energy of the iron edge in
the reflected component. The low ionization state can be realized for $\xi$ $<$ 1 in a photoionized plasma. 
The small $\xi$ suggests that iron emitting matter, considered to be reflected matter, is located at  $>$ 1 pc from the strong X-ray emitter.
This result is also deduced from the iron line width. The large distance  are consistent with  
a dusty torus.  The iron abundance was estimated by measuring the depth of the iron edge structure in the reflected component and 
by evaluating equivalent widths of a K$\alpha$ line from cold matter. Both results show that iron abundance is sub-solar 
assuming the Anders \& Grevesse abundances. Suzaku satellite can determine the EWs of emission lines with respect to 
the reflection component due to the Suzaku wide bandpass accurately.  

We detected a weak power-law emission component PL1 and many highly-ionized lines, which are associated with a highly-ionized plasma
which is photoionized by the inferred strong nuclear continuum. The weak power-law emission component was considered to be scattered light,  with a scattering
efficiency estimated to be about 1\%.  The efficiency corresponds to a value for the column density of the ionized matter of
about 3.6 $\times$10$^{22}$ cm$^{-2}$.  The ionized gas has a broad ionization structure, $\xi$ = 10--1000, deduced from the detection of
both He-like oxygen and He-like iron  lines.  The size of the scattering region was deduced to be about 50 pc, using a simple uniform density
model. However, to explain the spatial extent of the scattering region, the density may gradually decrease with as the distance from
the nucleus increases.

\vspace{1cm}

The authors wish to thank the staff of the Suzaku team. We also thank an anonymous referee
for useful comments. 
This study is carried out in part by the Grant-in-Aid for Scientific Research 
  (17030007 H.A. and 17740124 Y.T.) of the Ministry of Education, Culture, 
  Sports, Science and Technology.

\clearpage

%%%%%%%%%%%%%%%%%%%%%%%%%%%%%%%%%%%%%%%

%%%
% See the manual for the detail.
%%%

\clearpage

\begin{table}
\begin{center}
\caption{X-ray sources detected by Suzaku \label{tbl-1}}
\begin{tabular}{cccc}

\hline \hline

{ID} & {Position(J2000)} & {count rate ($\times$10$^{-6}$ ct/s/XIS)$^{*}$}  & {possible candidate} \\  \hline

1 & (06$^{\rm h}$ 13$^{\rm m}$ 43$^{\rm s}$ , $+$71$^{\circ}$ 07$^{\prime}$ 30$^{\prime\prime}$) & 450$\pm$10  &  87GB 060751.7+710809 (BL LAC) \\
2 & (06$^{\rm h}$ 13$^{\rm m}$ 46$^{\rm s}$ , $+$71$^{\circ}$ 03$^{\prime}$ 47$^{\prime\prime}$) & 59$\pm$5  & 1WGA J0613.7+7103 (ROSAT X-ray source)\\
3 & (06$^{\rm h}$ 14$^{\rm m}$ 53$^{\rm s}$ , $+$71$^{\circ}$ 05$^{\prime}$ 10$^{\prime\prime}$) & 50$\pm$5  & 2MASX J06145765+7105520 (?)\\
4 & (06$^{\rm h}$ 15$^{\rm m}$ 10$^{\rm s}$ , $+$71$^{\circ}$ 07$^{\prime}$ 34$^{\prime\prime}$) & 50$\pm$4  & UGC 03422 (?)\\
5 & (06$^{\rm h}$ 15$^{\rm m}$ 35$^{\rm s}$ , $+$71$^{\circ}$ 02$^{\prime}$ 19$^{\prime\prime}$) & 2200$\pm$24  & Mrk 3 \\
6 & (06$^{\rm h}$ 16$^{\rm m}$ 07$^{\rm s}$ , $+$71$^{\circ}$ 06$^{\prime}$ 38$^{\prime\prime}$) & 88$\pm$5  &  detected by XMM-Newton \\
7 & (06$^{\rm h}$ 16$^{\rm m}$ 16$^{\rm s}$ , $+$70$^{\circ}$ 56$^{\prime}$ 06$^{\prime\prime}$) & 20$\pm$4  & 1WGA J0616.2+7056 (ROSAT X-ray source)\\

\hline
\end{tabular}
\end{center}

*: Average count rates of 4 XIS detectors in the 0.4--2 keV band within a 1'.4 diameter circle after background subtraction.

\end{table}

\begin{table}
\begin{center}
\caption{Best-fit model parameters to the continuum emission \label{tbl-2}}
\begin{tabular}{cccccc}

\hline\hline
 photon index & $A_{\rm PL1}$ & $A_{\rm ref}$ & $N_{\rm H2}$ & $A_{\rm PL2}$ & $\chi^{2}$ (d.o.f.) \\
             &               &            &  (cm$^{-2}$) &   & \\
\hline
1.80$\pm$0.06 & (1.27$\pm$0.05)$\times$10$^{-4}$ & (1.56$\pm$0.11)$\times$10$^{-2}$ & (1.10$\pm$0.06)$\times$10$^{24}$  &  (1.15$\pm$0.20)$\times$10$^{-2}$ & 874.8 (762)\\
\hline
\end{tabular}
Note ---- $N_{\rm H1}$ was fixed at the Galactic column density.
\end{center}
\end{table}

\begin{table}
\begin{center}
\caption{Line energies \label{tbl-3}}
\tiny
\begin{tabular}{cccccccc}

\hline \hline

center energy$^{*}$               & line width & intensity                   & XMM-Newton RGS,pn$^{1}$ &   &  Chandra$^{2}$ & & ID \\
  (keV)    &  (eV)  &  ($\times$10$^{-6}$ ph cm$^{-2}$ s$^{-1}$)     &   (keV)        &   &    &  &  \\  \hline

0.495$^{+0.007}_{-0.006}$   & 1 (fixed)  &  33$\pm$7               & 0.50$^{+0.02}_{-0.03}$   & 14$^{+12}_{-4}$    &                 &          & N VII Ly$\alpha$ (0.5003 keV)\\
0.551$^{+0.003}_{-0.004}$   & 1 (fixed)  & 84$\pm$20                   & 0.562$\pm$0.001             & 63$^{+13}_{-22}$  & 0.561      & 37    & OVII K$\alpha$ f (0.561 keV) \\
0.566$^{+0.005}_{-0.007}$   & 1 (fixed)  & 52$\pm$14                   & 0.57$\pm$0.01                  & 25$^{+14}_{-13}$ & 0.569      & $<$13, $<$11   & OVII K$\alpha$ i+r (0.569/0.574 keV) \\
0.659$^{+0.003}_{-0.006}$   & 1 (fixed)  & 52$\pm$5                     & 0.655$\pm$0.005             & 32$^{+6}_{-8}$     & 0.654      & 11     & OVIII Ly$\alpha$ (0.654 keV) \\
0.736$\pm$0.003                    & 2 (fixed)  & 54$\pm$4                     & 0.741$^{+0.005}_{-0.008}$ & 28$^{+7}_{-11}$ & 0.780   & 6.3    & O VII RRC ( $>$0.739 keV) \\
0.816$^{+0.004}_{-0.005}$   & 2 (fixed)  & 36$\pm$3                     & 0.827$\pm$0.002            & 11$^{+7}_{-3}$      & 0.824      & 4.2    & Fe XVII L (0.826 keV), NeK? \\
0.870$\pm$0.009                    & 2 (fixed)  &  28$\pm$7                     & 0.88$^{+0.01}_{-0.02}$  & 16$\pm$4               & 0.873      & 3.9    & OVIII RRC ($>$0.871) \\
0.892$^{+0.009}_{-0.010}$   & 2 (fixed)  &   31$\pm$3                   & 0.91$^{+0.01}_{-0.03}$  & 18$^{+4}_{-8}$      & 0.905, 0.915 & $<$2.4, $<$2.5   & Ne IX K$\alpha$ f+i (0.905/0.915keV)  \\
0.920$\pm$0.003                    & 2 (fixed)  &    38$^{+5}_{-6}$          & 0.924$\pm$0.002            & 18$^{+4}_{-10}$    & 0.922     & 6.7     & Ne IX K$\alpha$ r (0.922keV) \\
1.006$\pm$0.003                     & 2 (fixed)  & 26.3$\pm$2.0              & 1.02$^{+0.05}_{-0.01}$  & 19$^{+4}_{-9}$   & 1.022 & 7.8     & Ne X Ly$\alpha$(1.022 keV) \\

1.066$\pm$0.004                    & 2 (fixed)  & 14.9$\pm$1.6              & 1.08$^{+0.01}_{-0.03}$  & 10$^{+8}_{-6}$   & 1.053  & 3.5     &  FeXXII L (1.053 keV)\\
1.152$\pm$0.006                    & 2 (fixed)  &  9.8$^{+1.4}_{-1.2}$   & 1.170                                  &  8$^{+8}_{-4}$   & 1.170  & 1.4  &  FeXXIII L (1.170 keV)\\                       
1.232$^{+0.008}_{-0.006}$   & 5 (fixed)  & 10.2$^{+1.3}_{-0.9}$  & 1.24$^{+0.02}_{-0.04}$  &  8$^{+7}_{-3}$   & 1.211 &  2    & Ne X Ly$\beta$ (1.211keV), Mg-K$\alpha$ (1.2536 keV) \\
1.337$\pm$0.004                    & 5 (fixed)  & 11.5$^{+1.1}_{-1.0}$  & 1.34$^{+0.01}_{-0.04}$  &  8$^{+7}_{-3}$   &  1.332 & $<$0.8 & Mg XI K$\alpha$ f(1.331keV) \\
                                                    &                  &                                        & 1.36$^{+0.03}_{-0.04}$  &  7$^{+7}_{-3}$   &  1.352 &  2.1   & Mg XI K$\alpha$ i+r (1.344/1.352 keV) \\
1.462$\pm$0.008                    & 5 (fixed)  & 4.5$\pm$0.9                & 1.47$^{+0.01}_{-0.02}$  &  8$^{+8}_{-4}$   &  1.473 &  1.9 & Mg XII Ly$\alpha$ (1.472 keV) \\
1.741$^{+0.006}_{-0.005}$   & 5 (fixed)  & 7.3$\pm$0.9                & 1.744$^{+0.018}_{-0.002}$$^{\dagger}$ & 7$\pm$1       &  1.7407$\pm$0.0073 & 2.2$\pm$0.4 & Si-K$\alpha$ (1.740 keV) \\
1.840$^{+0.013}_{-0.017}$   & 5 (fixed)  & 2.7$^{+1.1}_{-1.0}$    & 1.866$^{+0.012}_{-0.003}$$^{\dagger}$ & 6$\pm$1       & 1.841 &  2.8 & Si XIII f  (1.854 keV) \\
                                                    &                  &                                        &                                              &                  & 1.864 &  3.5 & Si XIII r+i (1.865keV) \\ 
1.998$\pm$0.013                     & 5 (fixed)  & 2.6$\pm$0.9                & 1.99$^{+0.03}_{-0.02}$$^{\dagger}$  &  2$\pm$1         & 2.008 &  3.5 & Si XIV Ly$\alpha$(2.005 keV) \\
2.315$\pm$0.011                    & 5 (fixed)  & 4.4$\pm$0.8                & 2.37$^{+0.04}_{-0.03}$$^{\dagger}$  &  4$\pm$1         &       &      & S-K$\alpha$ (2.307 keV) \\
2.444$^{+0.024}_{-0.030}$   & 5 (fixed)   & 1.4$^{+0.8}_{-0.9}$   &                                               &                   & 2.454 &  3.7 & S XV(2.447/2.461 keV) \\
2.612$^{+0.035}_{-0.040}$   & 5 (fixed)   & 1.0$^{+0.7}_{-0.8}$   &                                               &                   & 2.625 &  1.8 & S XVI Ly$\alpha$ (2.620 keV)\\
6.402$\pm$0.003                    & 20 ($<$ 29) & 47.4$^{+1.8}_{-2.0}$   &  6.415$\pm$0.006$^{\dagger}$    &   38$\pm$2    & 6.391$\pm$0.004 & 49$\pm$5  & Fe-K$\alpha$ (6.4038 keV) \\
6.682$^{+0.021}_{-0.023}$   & 0 ($<$42)  & 3.9$^{+1.5}_{-1.2}$  &  6.71$^{+0.03}_{-0.02}$$^{\dagger}$ &  4$\pm$2         & 6.68  &  21   & Fe XXV (6.637, 6.675, 6.700 keV) \\
7.090$^{+0.020}_{-0.022}$   & 20             & 4.9$^{+1.2}_{-1.4}$    &  7.06$^{\dagger}$      &  5$\pm$2  & 7.01      &  4.2  & H-like Fe (6.952/6.973 keV), Fe K$\beta$ 7.058 \\
7.424$^{+0.060}_{-0.050}$   & 20            & 3.8$\pm$1.1                &  7.60$\pm$0.05$^{\dagger}$           &  4$^{+1}_{-2}$   &       &      & Ni-K$\alpha$ (7.4782 keV) \\

\hline
\end{tabular}
\end{center}

$*$: Corrected for cosmological redshift (z=0.0135) \\
$\dagger$: XMM-Newton EPIC pn\\
1~Pounds, Page 2005, 2~Sako et al. 2000

\end{table}

\begin{table}
\begin{center}
\caption{Equivalent width \label{tbl-3}}
\tiny
\begin{tabular}{cccc}

\hline \hline

center energy$^{*}$               & EW$^{\dagger}$ to PL1 & EW$^{\dagger}$ to reflection  & ID       \\
  (keV)                            &  (eV)           &   (eV)                     &            \\  \hline

0.495$^{+0.007}_{-0.006}$   & 69 & 670      &  N VII Ly$\alpha$ (0.5003 keV)\\
0.551$^{+0.003}_{-0.004}$   & 290  & 6.7 k  &  OVII K$\alpha$ f (0.561 keV) \\
0.566$^{+0.005}_{-0.007}$   & 120  & 2.5 k  &  OVII K$\alpha$ i+r (0.569/0.574 keV) \\
0.659$^{+0.003}_{-0.006}$   & 190  & 2.9 k  &  OVIII Ly$\alpha$ (0.654 keV) \\
0.736$\pm$0.003                    & 240  & 3.4 k  &  O VII RRC ( $>$0.739 keV) \\
0.816$^{+0.004}_{-0.005}$   & 190  & 2.0 k  &  Fe XVII L (0.826 keV), NeK? \\
0.870$\pm$0.009                    & 280  & 3.0 k  &  OVIII RRC ($>$0.871) \\
0.892$^{+0.009}_{-0.010}$   & 120  & 1.2 k  &  Ne IX K$\alpha$ f+i (0.905/0.915keV)  \\
0.920$\pm$0.003                    & 350  & 3.4 k  &  Ne IX K$\alpha$ r (0.922keV) \\
1.006$\pm$0.003                    & 210  & 1.6 k  &  Ne X Ly$\alpha$(1.022 keV) \\

1.066$\pm$0.004                    & 130  & 900      &   FeXXII L (1.053 keV)\\
1.152$\pm$0.006                    &   99  &  540     &   FeXXIII L (1.170 keV)\\                       
1.232$^{+0.008}_{-0.006}$   & 130  & 580      &  Ne X Ly$\beta$ (1.211keV), Mg-K$\alpha$ (1.2536 keV) \\
1.337$\pm$0.004                    & 150  & 590      &  Mg IX K$\alpha$ f(1.331keV) \\
                                                    &                  &                                        &  Mg IX K$\alpha$ i+r (1.344/1.352 keV) \\
1.462$\pm$0.008                    &   68  & 210     &  Mg XII Ly$\alpha$ (1.472 keV) \\
1.741$^{+0.006}_{-0.005}$   & 150  & 300      &  Si-K$\alpha$ (1.740 keV) \\
1.840$^{+0.013}_{-0.017}$   &   62  & 120      &  Si XIII f  (1.854 keV),Si XIII r+i (1.865keV)\\
1.998$\pm$0.013                    &   68 & 100      & Si XIV Ly$\alpha$(2.005 keV) \\
2.315$\pm$0.011                    & 160  & 160      & S-K$\alpha$ (2.307 keV) \\
2.444$^{+0.024}_{-0.030}$   & 53   & 47 & S XV(2.447/2.461 keV) \\
2.612$^{+0.035}_{-0.040}$   & 41  & 32    &  S XVI Ly$\alpha$ (2.620 keV)\\
6.402$\pm$0.003                    & 10.8 k    & 855  & Fe-K$\alpha$ (6.4038 keV) \\
6.682$^{+0.021}_{-0.023}$   & 960        & 68 &  Fe XXV (6.637, 6.675, 6.700 keV) \\
7.090$^{+0.020}_{-0.022}$   & 1.30 k    & 81 &   H-like Fe (6.952/6.973 keV), Fe K$\beta$ 7.058 \\
7.424$^{+0.060}_{-0.050}$   & 1.1 k    & 120  &   Ni-K$\alpha$ (7.4782 keV) \\

\hline
\end{tabular}
\end{center}

*: corrected for cosmological redshift (z=0.0135) \\
$\dagger$: EW; equivalent width

\end{table}

\begin{table}
\begin{center}
\caption{Best-fit model parameters to the continuum emission with $plcabs$\label{tbl-4}}
\begin{tabular}{cccccc}

\hline\hline
  photon index & $A_{\rm PL1}$ & $A_{\rm ref}$ &  $N_{\rm H2}$ & $A_{\rm PL2}$ & $\chi^{2}$ (d.o.f.) \\
              &               &            &  (cm$^{-2}$) &   & \\
\hline
1.88$\pm$0.07 & (1.23$\pm$0.06)$\times$10$^{-4}$ & (1.86$\pm$0.10)$\times$10$^{-2}$ & (1.12$\pm$0.06)$\times$10$^{24}$  &  (1.69$\pm$0.17)$\times$10$^{-2}$ & 864.4 (762)\\
\hline
\end{tabular}
Note ----  $N_{\rm H1}$ was fixed at the Galactic column density.
\end{center}
\end{table}

\clearpage

%%%%%%%%%%%%%%%%%   Fig. 

\begin{figure}
\begin{center}

\FigureFile(100mm,50mm){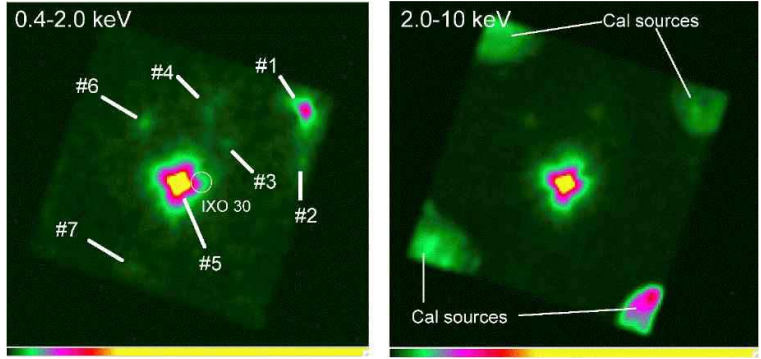}

\end{center}
\caption{X-ray images of the Mrk 3 region observed by Suzaku. The left and
right panels show the images in the 0.4--2 keV and 2--10 keV bands, respectively. }
\label{ptsrc_image}
\end{figure}%

%%%%%%%%%%%%%%%%%   Fig. 

\begin{figure}
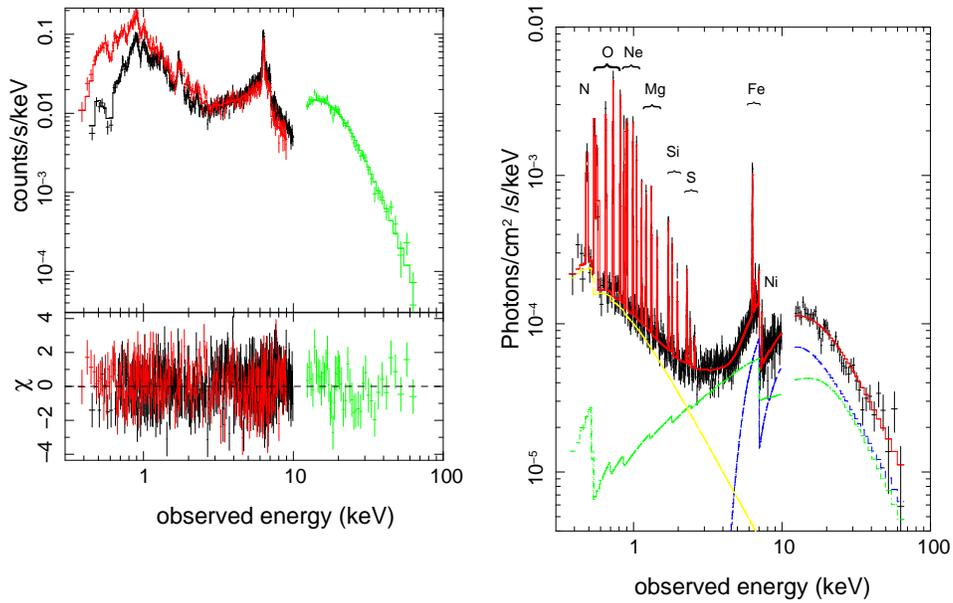

\begin{center}

\FigureFile(60mm,35mm){fig2a.ps}
\hspace*{2mm}
\FigureFile(60mm,30mm){fig2b.ps}

\end{center}
\caption{Wide-band spectrum of Mrk 3 observed with Suzaku (left). The black, red, and green crosses are
spectra obtained by the XIS-BI, XIS-FI, and HXD-PIN, respectively. The energy range around the Si-K edge (1.875--1.840 keV)
is ignored for the spectral fitting.  The right figure shows an unfolded X-ray spectrum. The yellow, green, and blue lines show the 
continuum emission: PL1, reflection, and heavily absorbed PL2, respectively. The red line shows the best-fit baseline model.}
\label{ptsrc_image}
\end{figure}%

%%%%%%%%%%%%%%%%%%  Fig. 
\begin{figure}
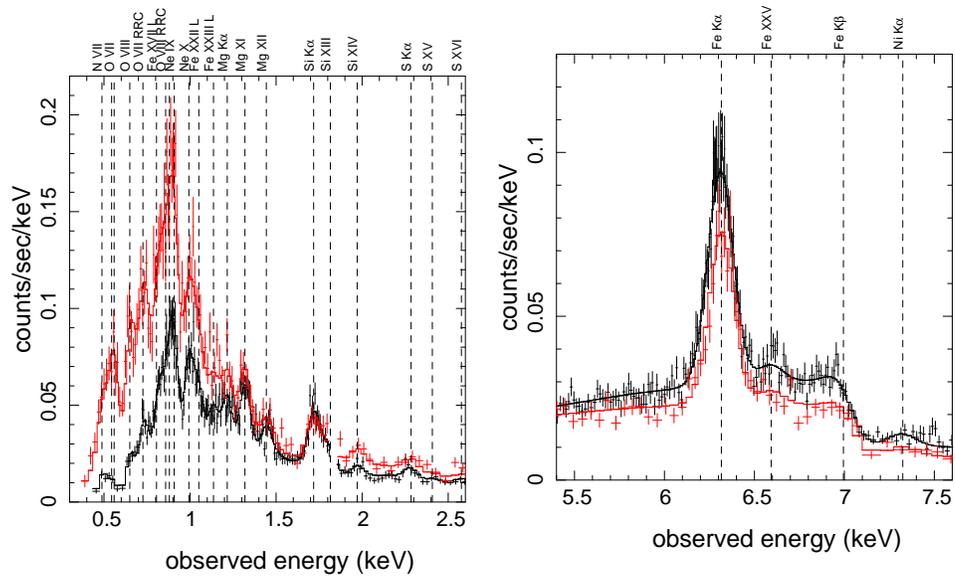

\begin{center}

\FigureFile(60mm,35mm){fig3a.ps}
\hspace*{2mm}
\FigureFile(60mm,30mm){fig3b.ps}

\end{center}
\caption{X-ray spectra in the 0.4--3keV and 5.5--7.5 keV bands. The best-fit center energies of the emission lines are denoted by 
the green lines.}
\label{ptsrc_image}
\end{figure}%

%%\begin{figure}
%%\begin{center}

%%\FigureFile(40mm,35mm){mrk3_soft_spec.ps}
%%\hspace*{1mm}
%%\FigureFile(40mm,35mm){mrk3_fe_spec.ps}

%%\end{center}
%%\caption{X-ray spectra in 0.4--3keV and 5.5--7.5 keV bands. }
%%\label{xray_spectrum2}
%%\end{figure}%

\begin{figure}
\begin{center}

\FigureFile(60mm,50mm){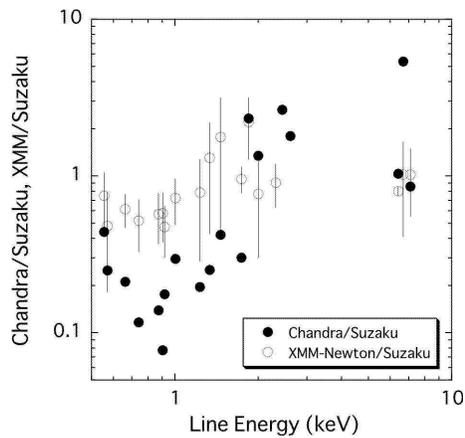}

\end{center}
\caption{Line intensity ratios between Suzaku and previous observations.}
\label{pin_lc}
\end{figure}%

%%%%%%%%%%%%%%%%%   Fig. 

\begin{figure}
\begin{center}

\FigureFile(60mm,35mm){fig5a.ps}
\hspace*{2mm}
\FigureFile(60mm,35mm){fig5b.ps}

\end{center}
\caption{The confidence contours of the fraction of Compton shoulder emission with respect to the K$\alpha$ core
versus the center energy (left) and the line width (right) of the iron line. The 68\%,
90\%, and 99\% confidence levels are plotted.}
\label{ptsrc_image}
\end{figure}%

%%%%%%%%%%%%%%%%%   Fig. 

\begin{figure}
\begin{center}

\FigureFile(60mm,35mm){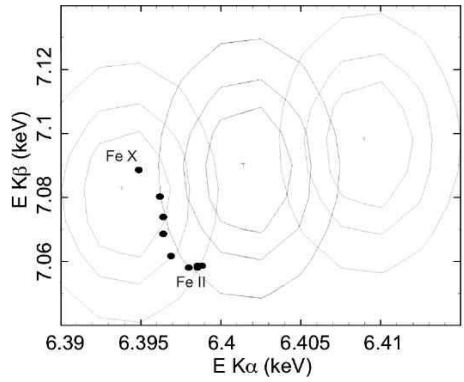}
%%\hspace*{2mm}
%%\FigureFile(60mm,35mm){ka-kb/I_Ka-kb.epsi}

\end{center}
\caption{The 68\%, 90\%, and 99\% confidence contours of  the center energies of iron K$\alpha$ versus K$\beta$. 
The filled circles denote the center energies of these lines for Fe II to Fe X (Palmeri et al. 2003).  Since the calibration 
uncertainty of the XIS response at 6 keV is about 15 eV, we show the contour maps with energy shifts of  +7.5 eV and -7.5 eV. 
%%The dependency of $\chi^{2}$ to an intensity ratio between iron K$\alpha$ and K$\beta$ (right).
}
\label{ptsrc_image}
\end{figure}%

%%%%%%%%%%%%%%%%%   Fig. 

\begin{figure}
\begin{center}

\FigureFile(60mm,30mm){fig7.ps}
%%\hspace*{2mm}
%%\FigureFile(60mm,35mm){reflection/K-edge_1.epsi}

\end{center}
\caption{The confidence contour of iron abundance determined by the edge feature seen in the reflection component and the absorption component. 
The 68\%,
90\%, and 99\% confidence levels are plotted.}
\label{ptsrc_image}
\end{figure}%

\begin{figure}
\begin{center}

\FigureFile(60mm,50mm){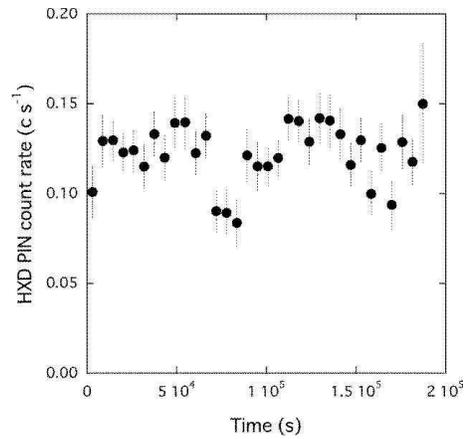}

\end{center}
\caption{The light curve of Mrk 3 in the 15--40 keV band. The error bars show only the statistical error of the count rates.
       A systematic error due to the non X-ray background subtraction is about 0.016 c s$^{-1}$.}
\label{pin_lc}
\end{figure}%

%%\begin{figure}
%%\begin{center}
%%
%%\FigureFile(60mm,50mm){ion-xi.epsi}
%%
%%\end{center}
%%\caption{Abundance of highly-ionized elements as a function of the ionization parameter $\xi$. The solid and dashed lines
%%corresponds to H-like and He-like ions, respectively.}
%%\label{pin_lc}
%%\end{figure}%

\end{document}